\documentclass{article}
\usepackage{graphicx}
\usepackage[preprint]{spconf}
\usepackage{amsmath,amssymb,bm,url}

\newcommand{\hyphen}{\mathchar`-}
\newcommand{\mathfont}{\mathsf}
\newcommand{\Freq}{F}
\newcommand{\freq}{f}
\newcommand{\minicirc}{\triangleleft}
\newcommand{\relu}{\mathfont{ReLU}}
\newcommand{\highway}{\mathfont{H}}
\newcommand{\trans}{^\mathfont{T}}
\newcommand{\dilate}{\delta}
\newcommand{\channels}{d}
\newcommand{\cnnbasic}[4]{#4\mathfont{C}{}^{{#1}}_{{#2} \star {#3}}}
\newcommand{\resizetensor}[4]{\cnnbasic{#1}{#2}{#3}{#4}}
\newcommand{\cnn}[4]{\cnnbasic{#1}{#2}{#3}{\highway}}
\newcommand{\dcnn}[4]{#4\mathfont{D}{{}^{{#1}}_{{#2} \star {#3}} } }
\newcommand{\att}[3]{\mathfont{Att}({#1},{#2},{#3})}

\newcommand{\tsume}{\vspace{-8pt}}

\usepackage{tikz}
\newcommand\copyrighttext{%
    \footnotesize \copyright 2018 IEEE. Personal use of this material is permitted.
    Permission from IEEE must be obtained for all other uses, in any current or future
    media, including reprinting/republishing this material for advertising or promotional
    purposes, creating new collective works, for resale or redistribution to servers or
    lists, or reuse of any copyrighted component of this work in other works.
}\newcommand\mycopyrightnotice{%
\begin{tikzpicture}[remember picture,overlay]
\node[anchor=north,yshift=-10pt] at 
    (current page.north) {\fbox{\parbox{\dimexpr\textwidth-\fboxsep-\fboxrule\relax}{\copyrighttext}}};
\end{tikzpicture}%
}

\title{Efficiently Trainable Text-to-Speech System Based on \\Deep Convolutional Networks With Guided Attention}
\twoauthors%
  {Hideyuki Tachibana,~~Katsuya Uenoyama}%
        {PKSHA Technology, Inc.\\Bunkyo, Tokyo, Japan\\%
            {\large\url{h_tachibana@pkshatech.com}}%
        }%
  {Shunsuke Aihara}%
        {Independent Researcher\\Shinjuku, Tokyo, Japan\\%
            {\large\url{aihara@argmax.jp}}%
        }%
\copyrightnotice{\copyright\ IEEE 2018.}
\toappear{Published as a conference paper at ICASSP 2018, Calgary, Canada}
\begin{document}
\ninept
\maketitle
\mycopyrightnotice
\begin{abstract}
This paper describes a novel text-to-speech (TTS) technique based on deep convolutional neural networks (CNN),
without use of any recurrent units.
Recurrent neural networks (RNN) have become a standard technique to model sequential data recently,
and this technique has been used in some cutting-edge neural TTS techniques.
However, training RNN components often requires a very powerful computer,
or a very long time, typically several days or weeks.
Recent other studies, on the other hand, have shown that CNN-based sequence synthesis can be much faster than RNN-based techniques,
because of high parallelizability.
The objective of this paper is to show that an alternative neural TTS based only on CNN
alleviate these economic costs of training.
In our experiment, the proposed Deep Convolutional TTS was sufficiently trained overnight (15 hours),
using an ordinary gaming PC equipped with two GPUs, while the quality of the synthesized speech was almost acceptable.
\end{abstract}
\begin{keywords}
Text-to-speech, deep learning, convolutional neural network, attention, sequence-to-sequence learning.
\end{keywords}
\section{Introduction}\label{sec:intro}
    Text-to-speech (TTS) is getting more and more common recently,
    and is getting to be a basic user interface for many systems.
    To further promote the use of TTS in various systems,
    it is significant to develop a manageable, maintainable, and extensible TTS component that is accessible to speech non-specialists,
    enterprising individuals and small teams.

    Traditional TTS systems, however, are not necessarily friendly for them,
    since they are typically composed of many domain-specific modules.
    For example, a typical parametric TTS system is an elaborate integration of many modules e.g.{} a text analyzer, an $F_0$ generator,
    a spectrum generator, a pause estimator, and a vocoder that synthesize a waveform from these data, etc.

    Deep learning~\cite{dlbook} may integrate these internal building blocks into a single model,
    and connects the input and the output directly. This type of technique is sometimes called `end-to-end' learning.
    Although such a technique is sometimes criticized as `a black box,'
    an end-to-end TTS system named Tacotron~\cite{Tacotron2017},
    which directly estimates a spectrogram from an input text,
    has achieved promising performance recently,
    without using hand-engineered parametric models based on domain-specific knowledge.
    Tacotron, however, has a drawback that it uses many recurrent units which are quite costly to train.
    It is almost infeasible for ordinary labs who do not have luxurious machines to study and extend it further.
    In fact, some people tried to implement open clones of Tacotron~\cite{tacotron:open1,tacotron:open2,tacotron:open3,tacotron:open4},
    but they are struggling to reproduce the speech of satisfactory quality as clear as the original results.

    The purpose of this paper is to show Deep Convolutional TTS (DCTTS), a novel fully convolutional neural TTS.
    The architecture is largely similar to Tacotron~\cite{Tacotron2017},
    but is based on a fully convolutional sequence-to-sequence learning model similar to the literature \cite{cs2s}.
    We show that this fully convolutional TTS actually works in a reasonable setting.
    The contribution of this article is twofold:
    (1) Propose a fully CNN-based TTS system
        which can be trained much faster than an
        RNN-based state-of-the-art neural TTS system, while the sound quality is still acceptable.
    (2) An idea to rapidly train the attention module, which we call `guided attention,' is also shown.
\subsection{Related Work}
\vspace{-5pt}
\subsubsection{Deep Learning and TTS}
Recently, deep learning-based TTS systems have been intensively studied,
and surprisingly high quality results are obtained in some of recent studies.
The TTS systems based on deep neural networks include Zen's work~\cite{zenicassp},
the studies based on RNN e.g.{}~\cite{rnntts1,rnntts2,rnntts3,rnntts4}, and recently proposed techniques e.g.{}
WaveNet~\cite[sec.~3.2]{wavenet}, Char2Wav~\cite{char2wav}, DeepVoice1\&2~\cite{deepvoice,deepvoice2},
and Tacotron~\cite{Tacotron2017}.

Some of them tried to reduce the dependency on hand-engineered internal modules.
The most extreme technique in this trend would be Tacotron~\cite{Tacotron2017},
which depends only on mel and linear spectrograms, and not on any other speech features e.g.~$F_0$.
Our method is close to Tacotron in a sense that it depends only on these spectral representations of audio signals.

Most of the existing methods above use RNN, a natural technique of time series prediction.
An exception is WaveNet, which is fully convolutional.
Our method is also based only on CNN but our usage of CNN would be different from WaveNet,
as WaveNet is a kind of a vocoder, or a back-end,
which synthesizes a waveform from some conditioning information that is given by front-end components.
On the other hand, ours is rather a front-end (and most of back-end processing).
We use CNN to synthesize a spectrogram, from which a simple vocoder can synthesize a waveform.

\vspace{-5pt}
\subsubsection{Sequence to Sequence $($seq2seq$)$ Learning}
    Recently, recurrent neural networks (RNN) have become a standard technique for mapping a sequence to another sequence,
    especially in the field of natural language processing, e.g.{} machine translation~\cite{cho,seq2seq}, dialogue system~\cite{dialoguele,dialogue}, etc.
    See also~\cite[sec.~10.4]{dlbook}.

    RNN-based seq2seq, however, has some disadvantages.
    Firstly, a vanilla encode-decoder model cannot encode too long sequence into a fixed-length vector effectively.
    This problem has been resolved by a mechanism called `attention'~\cite{nmt},
    and the attention mechanism now has become a standard idea of seq2seq learning techniques; see also~\cite[sec.~12.4.5.1]{dlbook}.

    Another problem is that RNN typically requires much time to train,
    since it is less suited for parallel computation using GPUs.
    To overcome this problem, several researchers proposed the use of CNN instead of RNN, e.g.~\cite{Kim2014,Zhang2015a,Kalchbrenner2016,Dauphin2016a,qrnn}.
    Some studies have shown that CNN-based alternative networks can be trained much faster,
    and sometimes can even outperform the RNN-based techniques.

    Gehring et~al.~\cite{cs2s} recently united these two improvements of seq2seq learning.
    They proposed an idea on how to use attention mechanism in a CNN-based seq2seq learning model,
    and showed that the method is quite effective for machine translation.
    The method we proposed is based on the similar idea to the literature~\cite{cs2s}.
\section{Preliminary}
\tsume
\subsection{Basic Knowledge of the Audio Spectrograms}
    An audio waveform and a complex spectrogram $Z = \{Z_{\freq t}\}\in \mathbb{C}^{\Freq' \times T'}$ are
    mutually transformed by linear maps called Short-term Fourier Transform (STFT) and inverse STFT,
    where  $\Freq'$ and $T'$ denote the number of frequency bins and temporal bins, respectively.
    It is common to consider only the magnitude $|Z| = \{|Z_{\freq t}|\} \in \mathbb{R}^{\Freq' \times T'}$,
    since it still has useful information for many purposes,
    and that $|Z|$ and $Z$ are almost the same in the sense that there exist many phase estimation
    (i.e.,{} $|Z|$ to $Z$ estimation, and therefore, waveform synthesis) techniques,
    e.g.{} the famous Griffin\&Lim algorithm (GLA)~\cite{gla}; see also e.g.~\cite{wileyphase}.
    We always use RTISI-LA~\cite{Zhu2007}, an online GLA, to synthesize a waveform.
    In this paper, we always normalize STFT spectrograms as $|Z| \gets (|Z|/\max(|Z| ))^\gamma$,
    and convert back $|Z| \gets |Z|^{\eta/\gamma}$ when we finally need to synthesize the waveform,
    where $\gamma, \eta$ are pre- and post-emphasis factors.

    It is also common to consider a mel spectrogram $S \in \mathbb{R}^{\Freq \times T'}$, $(\Freq \ll \Freq')$,
    obtained by applying a mel filter-bank to $|Z|$.
    This is a standard dimension reduction technique for speech processing.
    In this paper, we also reduce the temporal dimensionality from $T'$ to $\lceil T'/4 \rceil =: T$
    by picking up a time frame every four time frames,
    to accelerate the training of Text2Mel shown below.
    We also normalize mel spectrograms as $S \gets (S/\max(S))^\gamma$.

\subsection{Notation: Convolution and Highway Activation}\label{notation}
    In this paper, we denote the 1D convolution layer~\cite{convnet} by a space saving notation
    $\cnnbasic{o \gets i}{k}{\dilate}{} (X)$,
    where
    $ i $ is the size of input channel,
    $ o $ is the size of output channel,
    $ k $ is the size of kernel,
    $ \dilate $ is the dilation factor,
    and an argument $X$ is a tensor having three dimensions ({\it batch, channel, temporal}).
    The stride of convolution is always $1$.
    Convolution layers are preceded by appropriately-sized zero padding, whose size is suitably determined by a simple arithmetic
    so that the length of the sequence is kept constant.
    Let us similarly denote the 1D {\it deconvolution}
    layer as $\dcnn{o \gets i}{k}{\dilate}{} (X)$.
    The stride of deconvolution is always $2$ in this paper.
    Let us write a layer composition operator as $\cdot \minicirc \cdot$, and
    let us write networks like
    $\mathfont{F} \minicirc \mathfont{ReLU} \minicirc \mathfont{G} (X)
        := \mathfont{F}(\mathfont{ReLU} ( \mathfont{G} (X))),$
    and
    $(\mathfont{F} \minicirc \mathfont{G})^2(X)
        := \mathfont{F} \minicirc \mathfont{G} \minicirc \mathfont{F} \minicirc \mathfont{G} (X)$, etc.
    $\mathfont{ReLU}$ is an element-wise activation function defined by $\mathfont{ReLU}(x) = \max(x, 0)$.


    Convolution layers are sometimes followed by a Highway Network~\cite{highway}-like gated activation,
    which is advantageous in very deep networks:
    $
        \mathfont{Highway}(X; \mathfont{L}) = \sigma(H_1) \odot \mathfont{ReLU}(H_2) + (1 - \sigma(H_1)) \odot X,
    $
    where $H_1, H_2$ are tensors of the same shape as $X$, and are output as $[H_1, H_2]= \mathfont{L}(X)$ by a layer $\mathfont{L}$.
    The operator $\odot$ is the element-wise product, and $\sigma$ is the element-wise sigmoid function.
    Hereafter let us denote
    $
    \cnn{\channels \gets \channels}{k}{\dilate}{}(X) := \mathfont{Highway} (X; \cnnbasic{2d \gets d}{k}{\dilate}{})$.

\tsume
\section{Proposed Network}
\begin{figure}[!t]
    \centering
    \begin{minipage}[t]{0.95\linewidth}
          \centering
          \centerline{\includegraphics[width=0.85\columnwidth]{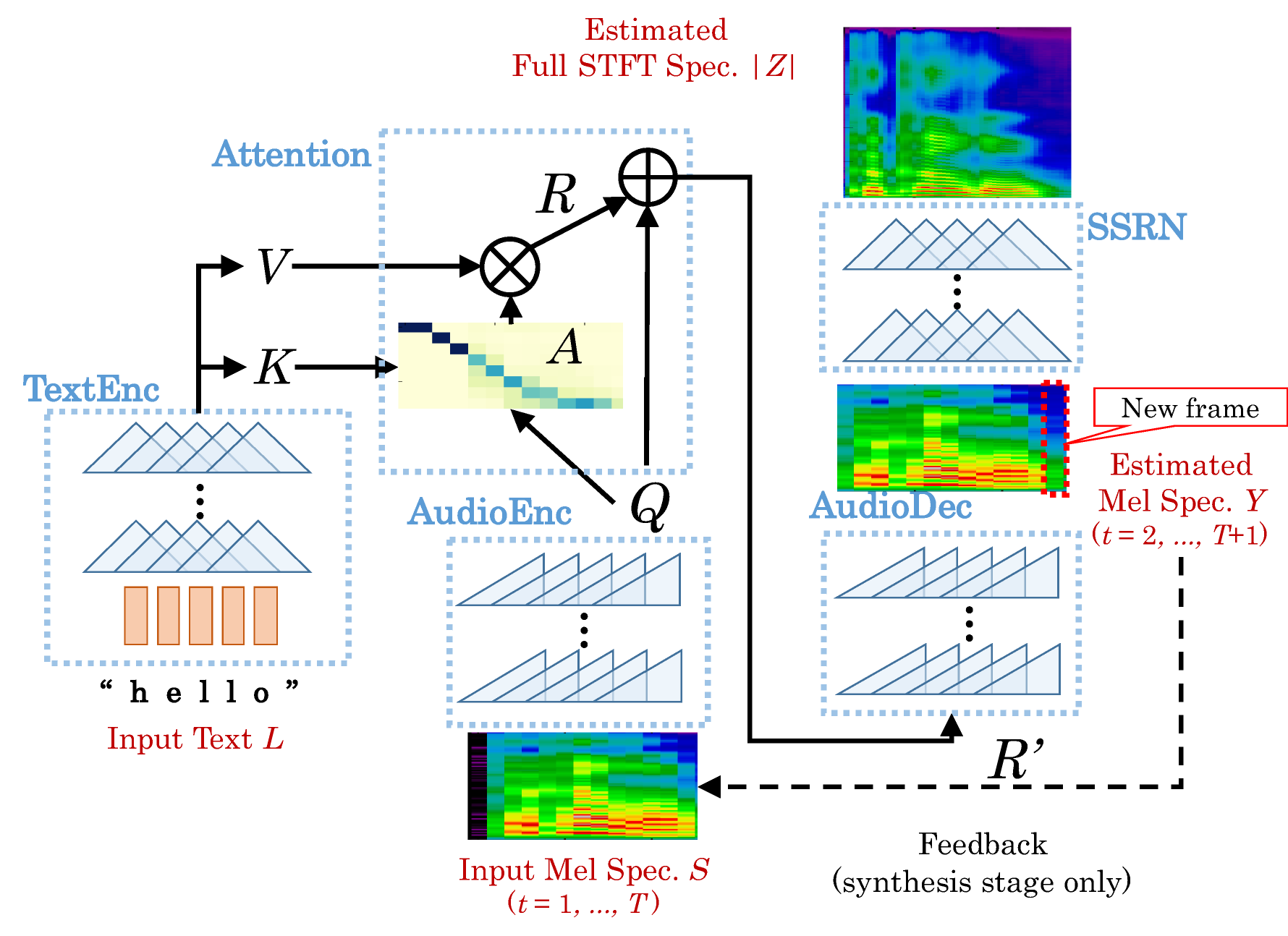}}
    \end{minipage}
    \vspace{-10pt}
    \caption{{ Network architecture.}}\label{fig:network}
    \vspace{-5pt}
    \begin{minipage}[t]{0.99\linewidth}
    {
    \footnotesize
    \hrulefill

    $
        \textstyle
            \mathsf{TextEnc}(L) :=
                (
                    \cnn{2d \gets 2d}{1}{1}{1}
                )^2 \minicirc
                (
                    \cnn{2d \gets 2d}{3}{1}{1}
                )^2 \minicirc
                (
                    \cnn{2d \gets 2d}{3}{27}{1} \minicirc
                    \cnn{2d \gets 2d}{3}{9}{1} \minicirc
                    \cnn{2d \gets 2d}{3}{3}{1} \minicirc
                    \cnn{2d \gets 2d}{3}{1}{1}
                )^2 \minicirc
                    \resizetensor{2d \gets 2d}{1}{1}{} \minicirc
                    \resizetensor{2d \gets e}{1}{1}{\relu\minicirc}\minicirc
                \mathsf{CharEmbed}^{e\hyphen\text{dim}}
            (L).\label{eq:textenc}
    $

        \hrulefill

    $
            \mathsf{AudioEnc}(S) :=
                (
                    \cnn{d \gets d}{3}{3}{1}
                )^2 \minicirc
                (
                    \cnn{d \gets d}{3}{27}{1} \minicirc
                    \cnn{d \gets d}{3}{9}{1} \minicirc
                    \cnn{d \gets d}{3}{3}{1} \minicirc
                    \cnn{d \gets d}{3}{1}{1}
                )^2 \minicirc
                    \resizetensor{d \gets d}{1}{1}{} \minicirc
                    \resizetensor{d \gets d}{1}{1}{\relu\minicirc} \minicirc
                    \resizetensor{d \gets \Freq}{1}{1}{\relu\minicirc}
            (S).\label{eq:audioenc}
    $

    \hrulefill

    $
            \mathsf{AudioDec}(R') :=
                \resizetensor{\Freq  \gets d}{1}{1}{\sigma\minicirc} \minicirc
                (
                    \cnnbasic{d \gets d}{1}{1}{\relu\minicirc}
                )^3 \minicirc
                (
                    \cnn{d \gets d}{3}{1}{1}
                )^2 \minicirc
                (
                    \cnn{d \gets d}{3}{27}{1} \minicirc
                    \cnn{d \gets d}{3}{9}{1} \minicirc
                    \cnn{d \gets d}{3}{3}{1} \minicirc
                    \cnn{d \gets d}{3}{1}{1}
                ) \minicirc
                \resizetensor{d \gets 2d}{1}{1}{}
            (R').\label{eq:audiodec}
    $

    \hrulefill

    $
            \mathsf{SSRN}(Y) :=
            \resizetensor{\Freq' \gets \Freq'}{1}{1}{\sigma\minicirc} \minicirc
            (
                \cnnbasic{\Freq' \gets \Freq'}{1}{1}{\relu \minicirc}
            )^2 \minicirc
            \resizetensor{\Freq' \gets 2c}{1}{1}{} \minicirc
            (
                \cnn{2c \gets 2c}{3}{1}{}
            )^2 \minicirc
            \resizetensor{2c \gets c}{1}{1}{}\minicirc
            (
                \cnn{c \gets c}{3}{3}{}\minicirc
                \cnn{c \gets c}{3}{1}{}\minicirc
                \dcnn{c \gets c}{2}{1}{}
            )^2\minicirc
            (
                \cnn{c \gets c}{3}{3}{}\minicirc
                \cnn{c \gets c}{3}{1}{}
            )\minicirc
        \resizetensor{c \gets \Freq}{1}{1}{}
        (Y).\label{eq:ssrn}
    $

    \hrulefill
    }

    \end{minipage}
    \vspace{-10pt}
    \caption{{ Details of each component. For notation, see section~\ref{notation}.}}\label{fig:network-detail}
\end{figure}
    Since some literature~\cite{Tacotron2017,stackgan} suggest that the staged synthesis from low- to high-resolution
    has advantages over the direct synthesis of high-resolution data,
    we synthesize the spectrograms using the following two networks.
    (1) Text2Mel, which synthesizes a mel spectrogram from an input text,
    and (2) Spectrogram Super-resolution Network (SSRN), which synthesizes a full STFT spectrogram from a coarse mel spectrogram.
    Fig.~\ref{fig:network} shows the overall architecture.

\subsection{Text2Mel: Text to Mel Spectrogram Network}
    We first consider to synthesize a coarse mel spectrogram from a text. This is the main part of the proposed method.
    This module consists of four submodules: Text Encoder, Audio Encoder, Attention, and Audio Decoder.
    The network $\mathfont{TextEnc}$ first encodes the input sentence
    $L = [l_1, \dots, l_N] \in \mathfont{Char}^N$
    consisting of $N$ characters, into the two matrices $K$ (key), $V$ (value) $\in \mathbb{R}^{d \times N}$, where $d$ the dimension of encoded characters.
    On the other hand, the network $\mathfont{AudioEnc}$ encodes the coarse mel spectrogram $S = S_{1:\Freq,1:T} \in \mathbb{R}^{\Freq \times T}$,
    of speech of length $T$,
    into a matrix $Q$ (query) $\in \mathbb{R}^{d \times T}$.
    \vspace{-2pt}%
    \begin{equation}
        (K, V) = \mathfont{TextEnc}(L), ~~
        Q = \mathfont{AudioEnc}(S_{1:\Freq,1:T}).
    \end{equation}%
    \vspace{-2pt}%
    An attention matrix $A \in \mathbb{R}^{N \times T}$, defined as follows,
    evaluates how strongly the $n$-th character $l_n$ and the $t$-th mel spectrum $S_{1:\Freq,t}$ are related.
    \vspace{-2pt}%
    \begin{equation}
        A = \mathfont{softmax}_{n\hyphen\text{axis}}(K\trans Q/\sqrt{d}).
    \end{equation}%
    \vspace{-2pt}%
    $A_{nt} \sim 1 $ implies that
    the module is focusing on $n$-th character $l_n$ at time $t$,
    and it will focus on $l_n$ or $l_{n+1}$ (or others nearby), at the subsequent time $t+1$.
    Whatever, let us expect those are encoded in $n$-th column of $V$.
    Thus a matrix $R \in \mathbb{R}^{d\times T}$, which is the `seed' of the subsequent mel spectra $S_{1:\Freq,2:T+1}$,
    is obtained as
    \vspace{-2pt}%
    \begin{equation}
    R = \att{Q}{K}{V} := V A. \text{~~~~(Note: matrix product.)}
    \end{equation}%
    \vspace{-2pt}%
    The resultant $R$ is concatenated with the encoded audio $Q$, as $R' = [R, Q]$,
    because we found it beneficial in our pilot study.
    Then, the Audio Decoder module estimates a mel spectrogram
    from the seed matrix $R' \in \mathbb{R}^{2d\times T}$ as follows,
    \vspace{-2pt}%
    \begin{equation}%
        Y_{1:\Freq,2: T+1} = \mathfont{AudioDec}(R').
    \end{equation}%
    \vspace{-2pt}%
    The resultant $Y_{1:\Freq, 2:T+1}$ needs to approximate the temporally-shifted ground truth $S_{1:\Freq, 2:T+1}$.
    The error is evaluated by a loss function $\mathcal{L}_\text{spec} (Y_{1:\Freq, 2:T+1} | S_{1:\Freq, 2:T+1} )$,
    and is back-propagated to the network parameters.
    The loss function was the sum of L1 loss and a function $\mathcal{D}_\text{bin}$ which we call binary divergence,
    \vspace{-7pt}%
    \begin{align}%
        \mathcal{D}_\text{bin} (Y|S)
        &:= \mathbb{E}_{\freq t} \left[
                - S_{\freq t}       \log \frac{Y_{\freq t}}{S_{\freq t}}
                - (1 - S_{\freq t}) \log \frac{1 - Y_{\freq t}}{1 - S_{\freq t}}
            \right] \nonumber \\
        &= \mathbb{E}_{\freq t} [- S_{\freq t} \hat{Y}_{\freq t} + \log(1 + \exp \hat{Y}_{\freq t}) ] + \mathrm{const.},\label{divergence}
    \end{align}
    \vspace{-2pt}%
        where $\hat{Y}_{\freq t} = \mathrm{logit}(Y_{\freq t})$.
        Since the gradient is non-vanishing, i.e.,
        $
        \partial \mathcal{D}_\text{bin} (Y|S)/\partial \hat{Y}_{\freq t} \propto Y_{\freq t} - S_{\freq t}
        $,
        it is advantageous for gradient-based training.
        It is easily verified that the spectrogram error is non-negative,
        $\mathcal{L}_\text{spec}(Y|S) = \mathcal{D}_\text{bin} (Y|S) + \mathbb{E}[|Y_{\freq t} - S_{\freq t}|] \ge 0$,
        and the equality holds iff $Y=S$.

\subsubsection{Details of TextEnc, AudioEnc, and AudioDec}
    The networks are fully convolutional, and are not dependent on any recurrent units.
    In order to take into account the long contextual information,
    we used {\it dilated convolution}~\cite{dilatedcnn,wavenet,Kalchbrenner2016}
    instead of RNN.

    The top equation of Fig.~\ref{fig:network-detail} is the content of $\mathsf{TextEnc}$.
    It consists of a character embedding and several 1D {\it non-causal} convolution layers.
    In the literature~\cite{Tacotron2017} an RNN-based component named `CBHG' was used, but we found this convolutional network also works well.
    $\mathsf{AudioEnc}$ and $\mathsf{AudioDec}$, shown in Fig.~\ref{fig:network-detail}, are composed of 1D {\it causal} convolution layers.
    These convolution should be causal because the output of $\mathsf{AudioDec}$ is feedbacked to the input of $\mathsf{AudioEnc}$ at the synthesis stage.

\subsection{Spectrogram Super-resolution Network (SSRN)}

    We finally synthesize a full spectrogram $|Z| \in \mathbb{R}^{\Freq' \times 4T}$,
    from the obtained coarse mel spectrogram $Y \in \mathbb{R}^{\Freq \times T}$,
    using the spectrogram super-resolution network (SSRN).
    Upsampling frequency from $\Freq$ to $\Freq'$ is fairly simple.
    We can achieve that by increasing the convolution channels of a 1D convolutional network.
    Upsampling of the temporal axis is not done the same way.
    Instead,
    we quadruple the length of the sequence from $T$ to $4T=T'$,
    by applying `deconvolution' layers of stride size 2 twice.

    The bottom equation of Fig.~\ref{fig:network-detail} shows $\mathsf{SSRN}$.
    In this paper, all convolutions of SSRN are non-causal, since we do not consider online processing.
    The loss function is the same as Text2Mel: the sum of  L1 distance and $\mathcal{D}_\text{bin}$, defined by \eqref{divergence},
    between the synthesized spectrogram $\mathsf{SSRN}(S)$ and the ground truth $|Z|$.

\tsume
\section{Guided Attention}
\tsume
\subsection{Guided Attention Loss: Motivation, Method and Effects}
    In general, an attention module is quite costly to train.
    Therefore, if there is some prior knowledge, incorporating them into the model may be a help to alleviate the heavy training.
    We show that the following simple method helps to train the attention module.

    In TTS, the possible attention matrix $A$ lies in the very small subspace of $\mathbb{R}^{N\times T}$.
    This is because of the rough correspondence of the order of the characters and the audio segments.
    That is, when one reads a text, it is natural to assume that the text position $n$ progresses nearly linearly to the time $t$,
    i.e., $n \sim a t$, where $a \sim N/T$.
    This is a noticeable difference of TTS from other seq2seq learning techniques such as machine translation,
    where an attention module needs to resolve the word alignment between two languages that have very different syntax, e.g.{} English and Japanese.

    Based on this idea, we introduce `guided attention loss,' which prompts the attention matrix $A$ to be `nearly diagonal,'
    $
        \mathcal{L}_\text{att}(A) = \mathbb{E}_{nt}[ A_{nt} W_{nt} ],
    $
    where
    $W_{nt} = 1 - \exp\{-{(n/N - t/T)^2}/{2g^2}\}.$
    In this paper, we set $g=0.2$.
    If $A$ is far from diagonal (e.g., reading the characters in the random order),
    it is strongly penalized by the loss function.
    This subsidiary loss function is simultaneously optimized with the main loss $\mathcal{L}_\text{spec}$ with the equal weight.

\begin{figure}[!t]
    \vspace{-10pt}
    \centering
    \begin{minipage}[t]{0.99\linewidth}
        \centering
        \centerline{\includegraphics[width=0.99\columnwidth]{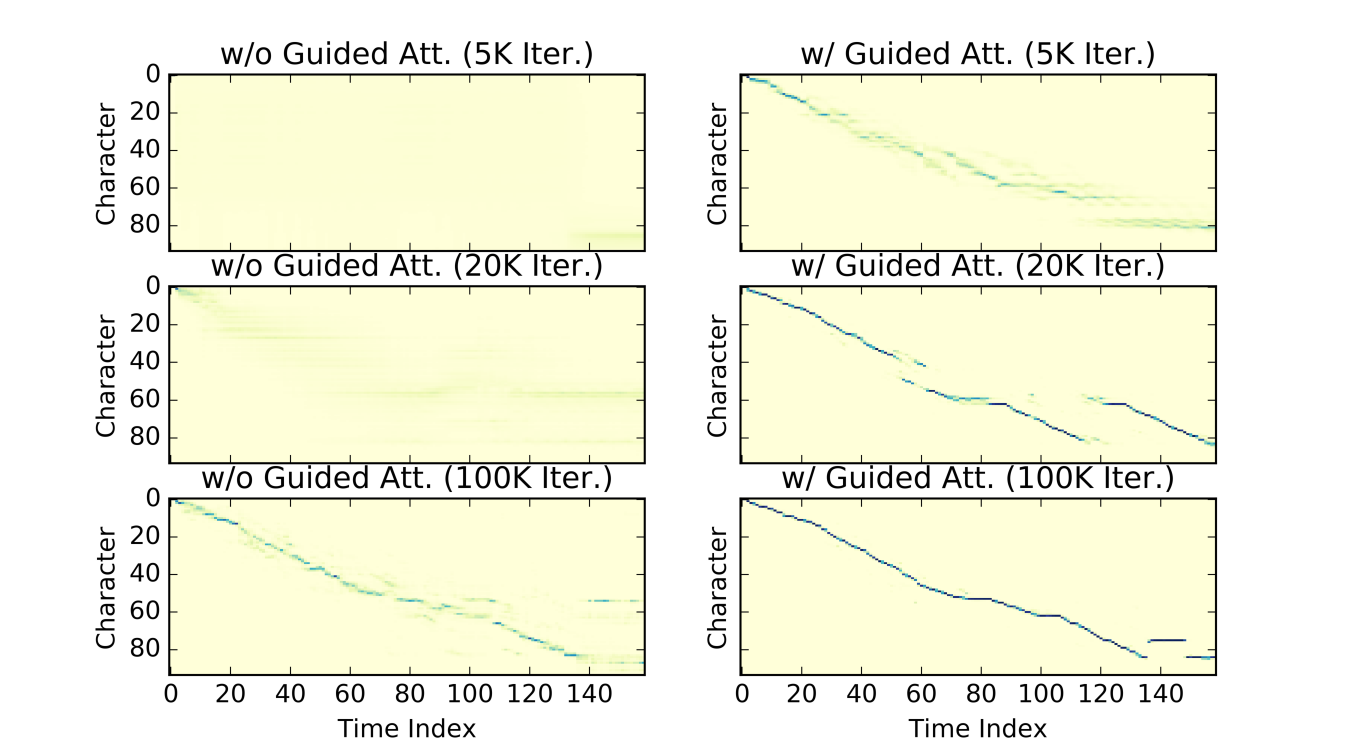}}
        \vspace{-10pt}
        \caption{
            {
                Comparison of the attention matrix $A$, trained with and without the guided attention loss $\mathcal{L}_\text{att}(A)$.
                (Left) Without, and (Right) with the guided attention.
                The test text is {\it ``icassp stands for the international conference on acoustics, speech, and signal processing."}
                We did not use the heuristics described in section~\ref{sec:fia}.
            }
        }
        \label{fig:attention}
    \end{minipage}
    \vspace{-10pt}
\end{figure}

    Although this measure is based on quite a rough assumption, it improved the training efficiency.
    In our experiment, if we added the guided attention loss to the objective,
    the term began decreasing only after $\sim$100 iterations.
    After $\sim$5K iterations, the attention became roughly correct, not only for training data, but also for new input texts.
    On the other hand, without the guided attention loss, it required much more iterations.
    It began learning after $\sim$10K iterations,
    and it required $\sim$50K iterations to look at roughly correct positions, but the attention matrix was still vague.
    Fig.~\ref{fig:attention} compares the attention matrix, trained with and without guided attention loss.
\tsume
\subsection{Forcibly Incremental Attention at the Synthesis Stage}\label{sec:fia}
    At the synthesis stage, the attention matrix $A$ sometimes fails to focus on the correct characters.
    Typical errors we observed were (1) skipping several letters, and (2) repeating a same word twice or more.
    In order to make the system more robust,
    we heuristically modified the matrix $A$ to be `nearly diagonal,' by a simple rule as follows.
    We observed this method sometimes alleviated such failures.

    {\it
        Let $n_t$ be the position of the character which is read at time $t$, i.e., $n_t = \mathrm{argmax}_n A_{nt}$.
        Comparing the current position $n_t$ and the previous position $n_{t-1}$,
        if the difference $n_t - n_{t-1}$ is outside of the range
        $-1 \le n_t - n_{t-1} \le 3$, the current attention is forcibly set as
        $A_{nt} = \delta_{n, n_{t-1}+1}$ $($Kronecker delta$)$,
        to increment the attention target as $n_t = n_{t-1} + 1$.
    }

\tsume
\section{Experiment}
\tsume
\subsection{Experimental Setup}
\begin{table}[t]
    \centering
    \caption{{ Parameter Settings.}}
    \label{tab:params}
    {
        \footnotesize
        \begin{tabular}{l|l} \hline
            Sampling rate of audio signals & 22050 Hz\\ \hline
            STFT window function& Hanning\\ \hline
            STFT window length and shift & 1024 ($\sim$46.4 [ms]), 256 ($\sim$11.6[ms])\\ \hline
            STFT spectrogram size $\Freq' \times 4T$ & $513 \times 4T$ ($T$ depends on audio clip) \\ \hline
            Mel spectrogram size $\Freq \times T$ & $80 \times T$ ($T$ depends on audio clip) \\ \hline
            Dimension $e$, $d$ and $c$ & 128, 256, 512 \\ \hline
            ADAM parameters $(\alpha, \beta_1, \beta_2, \varepsilon)$ & $(2 \times 10^{-4}, 0.5, 0.9, 10^{-6})$ \\ \hline
            Minibatch size & 16 \\ \hline
            Emphasis factors $(\gamma, \eta)$ & (0.6, 1.3) \\ \hline
            RTISI-LA window and iteration & 100, 10 \\ \hline
            Character set, $\mathsf{Char}$ & \texttt{a-z,.'-} and \texttt{Space} and \texttt{NULL} \\ \hline
        \end{tabular}
    }
    \vspace{-7pt}
    \centering
        \caption{{ Comparison of MOS (95\% confidence interval), training time, and iterations (Text2Mel/SSRN),
        of an open Tacotron~\cite{tacotron:open3} and the proposed method (DCTTS).
        The digits with * were excerpted from the repository~\cite{tacotron:open3}}.
        }
        \label{tab:comparison}
        {
            \footnotesize
            \begin{tabular}{l|rr|r} \hline
            Method & Iteration & Time & MOS (95\% CI) \\ \hline \hline
            Open Tacotron~\cite{tacotron:open3} &
                        877K${}^*$ & 12 days${}^*$  & $2.07 \pm 0.62$ \\ \hline
            DCTTS
                        &  20K/\phantom{9}40K & $\sim$2 hours & $1.74 \pm 0.52$ \\ \hline
            DCTTS
                        &  90K/150K & $\sim$7 hours & $2.63 \pm 0.75$ \\ \hline
            DCTTS
                        & 200K/340K & $\sim$15 hours & $2.71 \pm 0.66$ \\ \hline
            DCTTS
                        & 540K/900K & $\sim$40 hours & $2.61 \pm 0.62$ \\ \hline
            \end{tabular}
        }
        \vspace{-5pt}
\end{table}

    We used LJ Speech Dataset ~\cite{ljspeech} to train the networks.
    This is a public domain speech dataset
    consisting of $\sim$13K pairs of text and speech, without phoneme-level alignment, $\sim$24 hours in total.
    These speech data have a~little reverberation.
    We preprocessed the texts by spelling out some of abbreviations and numeric expressions,
    decapitalizing the capitals, and removing less frequent characters not shown in Table~\ref{tab:params},
    where \texttt{NULL} is a dummy character for zero-padding.

    We implemented our neural networks using Chainer~2.0~\cite{chainer}.
    We trained the models using a household gaming PC equipped with two GPUs.
    The main memory of the machine was 62GB, which is much larger than the audio dataset.
    Both GPUs were NVIDIA GeForce GTX 980 Ti, with 6 GB memories.

    For simplicity, we trained Text2Mel and SSRN independently and asynchronously using different GPUs.
    All network parameters were initialized using He's Gaussian initializer~\cite{henorm}.
    Both networks were trained by the ADAM optimizer~\cite{adam}.
    When training SSRN, we randomly extracted short sequences of length $T=64$ for each iteration to save memory usage.
    To reduce the disk access, we reduced the frequency of creating the snapshot of parameters to only once per 5K iterations.
    Other parameters are shown in Table~\ref{tab:params}.

    As it is not easy for us to reproduce the original results of Tacotron,
    we instead used a ready-to-use model~\cite{tacotron:open3} for comparison,
    which seemed to produce the most reasonable sounds in the open implementations.
    It is reported that this model was trained using LJ Dataset
    for 12 days (877K iterations) on a GTX 1080 Ti, newer GPU than ours.
    Note, this iteration is still much less than the original Tacotron, which was trained for more than 2M iterations.

    We evaluated mean opinion scores (MOS) of both methods by crowdsourcing on Amazon Mechanical Turk using crowdMOS toolkit~\cite{crowdmos}.
    We used 20 sentences from {\it Harvard Sentences} List 1\&2.
    The audio data were synthesized using five methods shown in Table~\ref{tab:comparison}.
    The crowdworkers rated these 100 clips from 1 (Bad) to 5 (Excellent).
    Each worker is supposed to rate at least 10 clips.
    To obtain more responses of higher quality, we set a few incentives shown in the literature.
    The results were statistically processed using the method shown in the literature~\cite{crowdmos}.

\subsection{Result and Discussion}
\begin{figure}[t]
    \begin{minipage}[t]{0.99\linewidth}
          \centering
            \vspace{-5pt}
          \centerline{\includegraphics[width=0.99\columnwidth]{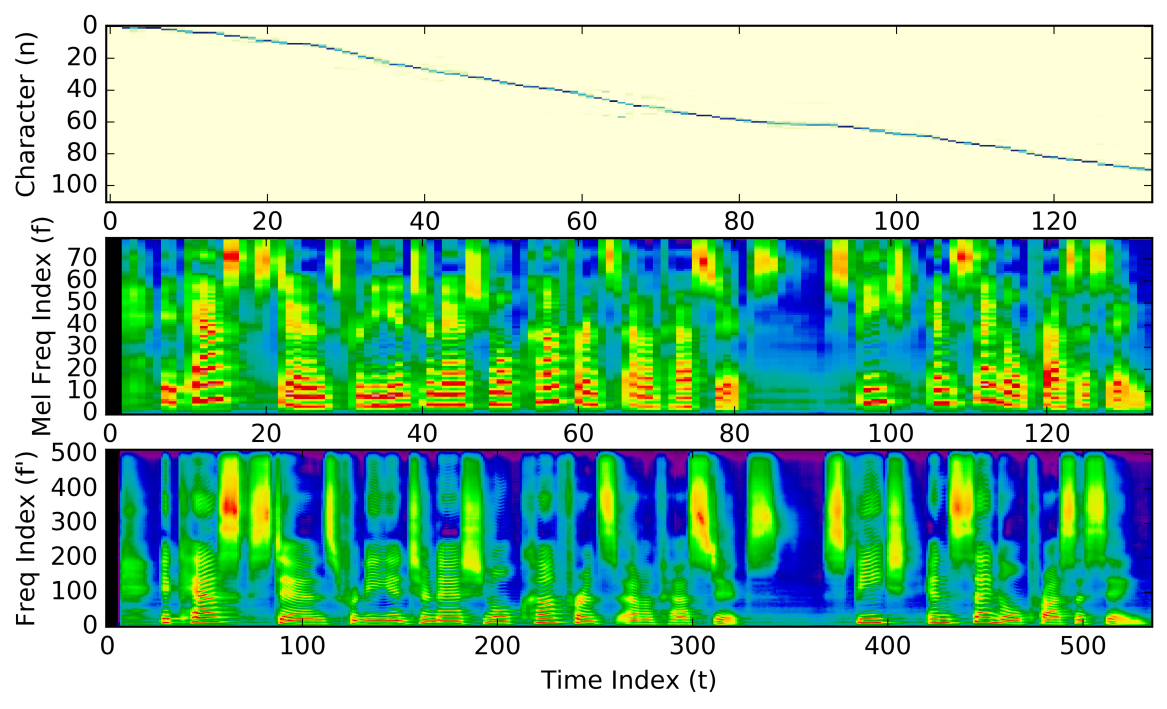}}
          \vspace{-15pt}
          \caption{
            {
                (Top) Attention, (middle) mel, and (bottom) linear STFT spectrogram, synthesized by the proposed method, after 15 hours training.
                The input text is {\it``icassp stands for the international conference on acoustics, speech and signal processing."} (90 chars)
          }}\label{fig:result}
    \end{minipage}
    \vspace{-5pt}
\end{figure}

    In our setting, the training throughput was $\sim$3.8 minibatch/s (Text2Mel) and $\sim$6.4 minibatch/s (SSRN).
    This implies that we can iterate the updating formulae of Text2Mel 200K times in 15 hours.
    Fig.~\ref{fig:result} shows an example of attention, synthesized mel and full spectrograms, after 15 hours training.
    It shows that the method can almost correctly focus on the correct characters, and synthesize quite clear spectrograms.
    More samples are available at the author's web page.\footnote{\url{https://github.com/tachi-hi/tts_samples}}

    In our crowdsourcing experiment, 31 subjects evaluated our data.
    After the automatic screening by the toolkit~\cite{crowdmos}, 560 scores rated by 6 subjects were selected for final statistics calculation.
    Table~\ref{tab:comparison} compares the performance of our proposed method (DCTTS) and an open Tacotron.
    Our MOS (95\% confidence interval) was $2.71 \pm 0.66$ (15 hours training) while the Tacotron's was $2.07 \pm 0.62$.
    Although it is not a strict comparison since the frameworks and the machines were different,
    it would be still concluded that our proposed method is quite rapidly trained to the satisfactory level compared to Tacotron.

    Note that the MOS were below the level reported in~\cite{Tacotron2017}.
    The reasons may be threefold: (1) the limited number of iterations,
    (2) SSRN needs to synthesize the spectrograms from less information than~\cite{Tacotron2017},
    and (3) the reverberation of the training data.

\tsume
\section{Summary and Future Work}
    This paper described a novel TTS technique based on deep convolutional neural networks,
    and a technique to train the attention module rapidly.
    In our experiment, the proposed Deep Convolutional TTS was trained overnight ($\sim$15 hours),
    using an ordinary gaming PC equipped with two GPUs, while the quality of the synthesized speech was almost acceptable.

    Although the audio quality is far from perfect yet,
    it may be improved
    by tuning some hyper-parameters thoroughly, and
    by applying some techniques developed in the deep learning community.

    We believe this method will encourage further development of the applications based on speech synthesis.
    We can expect that this simple neural TTS may be extended to many other purposes
    e.g.{} emotional/non-linguistic/personalized speech synthesis, etc.,
    by further studies.
    In addition, since a neural TTS has become lighter, the studies on more integrated speech systems
    e.g.{} some multimodal systems,
    may have become more feasible.
    These issues should be worked out in the future.

\tsume
\section{Acknowledgement}
    The authors would like to thank the OSS contributors
    and the data creators (LibriVox contributors and @keithito).

\bibliographystyle{IEEEbib}
{
    \bibliography{mybib}
}
\end{document}